\documentclass[pra,aps,twocolumn,superscriptaddress,showpacs,floatfix,letterpaper]{revtex4-1}  
\usepackage{hyperref}
\usepackage{amssymb}
\usepackage{amsmath}
\usepackage{bbm}
\usepackage{graphicx}
\usepackage{epsfig}
\usepackage{epstopdf}
\usepackage[usenames]{color}
\usepackage{soul}

\hypersetup{
  colorlinks = true, 
  urlcolor   = blue, 
  linkcolor  = blue, 
  citecolor  = blue  
}

\begin{document}
\bibliographystyle{apsrev4-1}

\title{Ferromagnetism and correlation strength in cubic barium ruthenate in comparison to strontium and calcium ruthenate: a dynamical mean field study}

\author{Qiang Han}
\affiliation{Department of Physics, Columbia University, New York, New  York 10027, USA}

\author{Hung T. Dang}
\affiliation{Institute for Theoretical Solid State Physics, JARA-FIT and JARA-HPC, RWTH Aachen University, 52056 Aachen, Germany}

\author{A. J. Millis }
\affiliation{Department of Physics, Columbia University, New York, New  York 10027, USA}

\date{\today}

\begin{abstract}
We present density functional plus dynamical mean field studies of cubic BaRuO$_3$ using interaction parameters previously found to be appropriate for the related materials CaRuO$_3$ and SrRuO$_3$. The calculated variation in transition temperature between the Ba and Sr compounds is consistent with experiment, confirming the assignment of the compounds to the Hund's metal family of materials, and also confirming the appropriateness of the values for the interaction parameters previously estimated and the appropriateness of the single-site dynamical mean field approximation for these materials.  The results provide insights into the origin of magnetism and the role of the van Hove singularity in the physics of Hund's metals. 
\end{abstract}

\pacs{71.27.+a,75.50.Cc,72.15.Eb}

\maketitle

\section{Introduction}

The relation between crystal structure and electronic properties is a fundamental issue in condensed matter and materials physics. Studies of the variation of properties across a family of materials with similar chemical composition and structures can provide insight while the ability to capture the variation in properties is an important  test of theoretical methods. In this paper we consider the $A$RuO$_3$ pseudocubic ruthenium-based perovskite family of materials, with $A$=Ca, Sr or Ba. The Sr and Ca compounds have been studied for decades, but BaRuO$_3$ has been successfully synthesized only relatively recently~\cite{Jin20052008,PhysRevLett.101.077206}. The materials crystallize in variants of the ideal $AB$O$_3$ cubic perovskite structure and the three choices of $A$ site ion are 'isoelectronic': each donates two electrons to the Ru-O complex and is otherwise electrically inert at the relevant energy scales. All three compounds display correlated electron behavior including large mass enhancements. the Ca material is paramagnetic down to the lowest temperatures measured, while the   Sr and Ba materials have ferromagnetic ground states with the transition temperature of the Sr materials rather higher than that of the Ba material. The Ba compound is cubic; in the Sr and Ca materials a GdFeO$_3$ distortion (rotation and tilt of the RuO$_6$ octahedra) occurs, with the rotation and tilting angles being larger in the Ca than in the Sr compound. In the Ba compound a van Hove singularity leads to a density of states peak that happens to be very close to the Fermi level. The GdFeO$_3$ distortion splits and weakens the peak in the Ca and Sr materials; thus comparison of the electronic properties provides insight into the role of the van Hove singularity in the magnetic ordering and correlation physics.

In this paper we present a comparative density functional plus dynamical mean field (DFT+DMFT) analysis of Ba-, Sr- and CaRuO$_3$ aimed at gaining understanding of the relation between the degree of octahedral distortion, the correlation strength, and the magnetism in this family of compounds. Our work builds on a  DFT+DMFT study of CaRuO$_3$ and SrRuO$_3$ by Mravlje, Georges and two of us ~\cite{PhysRevB.91.195149} which  concluded that the pseudocubic ruthenates should be identified as `Hund's metals' in which the physics is dominated by a slowly fluctuating local moment in the Ru d-shells while Mott physics is of secondary importance~\cite{Georges13}.  A particular combination of interaction parameters was argued to describe the materials well. In this paper we use the same methods to calculate the ferromagnetic transition temperature and electron self energy for cubic BaRuO$_3$, fixing the interaction parameters to the values determined previously. We find that the calculated difference in ferromagnetic transition temperature between the Sr and Ba compounds is in good accord with experiment, confirming both the applicability of the density functional plus dynamical mean field methodology to these compounds  and the correctness of the interaction parameters. Consideration of the variation of the  electron self energy across the series of compounds is shown to  lead to  insight into the role of the van Hove singularity in the physics of Hund's metals.  DFT+DMFT methods have been used to study BaRuO$_3$ \cite{PhysRevB.87.165139,arXiv:1511.01371}, but the comparative study we present of the three ruthenate materials is new. We will comment on the relation between our work and that of Refs.~\onlinecite{PhysRevB.87.165139,arXiv:1511.01371} below.

The rest of this paper is organized as follows. Section~\ref{sec:formalism} presents the calculational methodology. Section~\ref{sec:results} presents our main calculated results and section~\ref{sec:analysis} provides analysis and interpretation of the transition temperatures. Section~\ref{sec:correlation} discusses the issue of the relative correlation strengths of the materials. Section~\ref{sec:conclusion} is a summary and conclusion. 
 
\section{Crystal structure, electronic structure and many-body model\label{sec:formalism}}

\subsection{Crystal Structures}
BaRuO$_3$ crystallizes in the ideal $AB$O$_3$ perovskite structure with bond length $2.003$~\AA~\cite{Jin20052008}. CaRuO$_3$ and SrRuO$_3$ crystallize in a $Pnma$ symmetry crystal structure related to the ideal cubic perovskite structure by a GdFeO$_3$ distortion corresponding to a four-sublattice tilt and rotation of the RuO$_6$ octahedra. The Ru-O-Ru bond angles of the three compounds are $180^\circ$ (BaRuO$_3$), $\approx163^\circ$ (SrRuO$_3$) and   $\approx 150^\circ$ (CaRuO$_3$)  \cite{Jones89,Bensch90}.

\subsection{Background electronic structure}

\begin{figure}[t]
\centering
\includegraphics[width=\columnwidth]{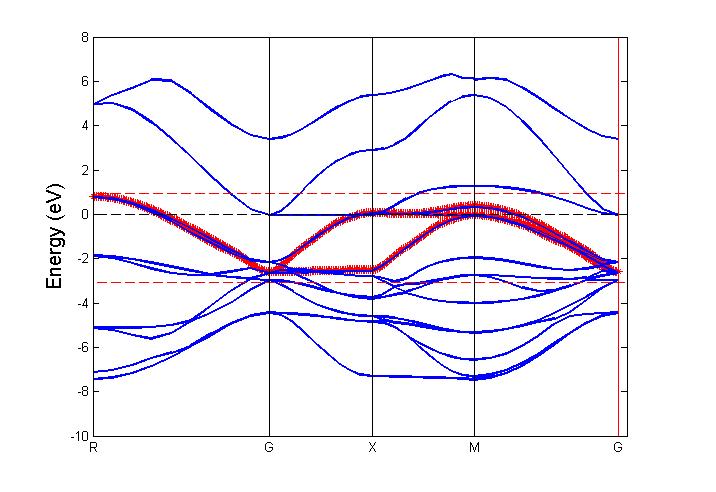}\\(a)\\
\includegraphics[width=\columnwidth]{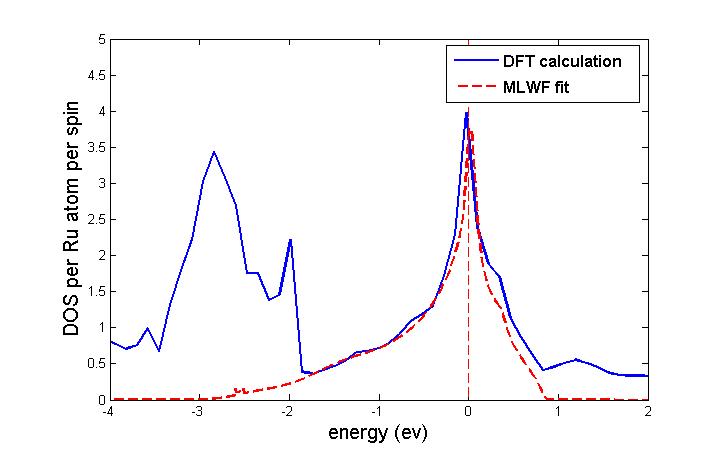}\\(b)
\caption{(Color online) (a) Near Fermi level energy bands  of cubic perovskite BaRuO$_3$. Lighter lines (blue on-line) are DFT bands. Heavier lines (red online) are MLWF fits to the $t_{2g}$-derived near Fermi level orbitals using an energy window extending from $-3$~eV to $1$~eV (dashed red line). (b) Total density of states per Ru atom for BaRuO$_3$:  solid lines (blue online) DFT results; dashed lines (red online) MLWF fit. The Fermi level is at energy $\omega=0$.}
\label{fig:BaRuO3bands}
\end{figure}

We computed band structures for CaRuO$_3$, SrRuO$_3$  and BaRuO$_3$ using the experimental atomic positions and the non-spin-polarized Generalized Gradient Approximation as implemented in VASP~\cite{Kresse93,Kresse96a,Kresse96b,Kresse99}  with energy cutoff 400~eV and k-mesh as large as $11\times11\times11$ to verify convergence. (Figure~\ref{fig:BaRuO3bands} is based on this mesh. The rest of our results are obtained from a k-mesh of $5\times5\times5$ to obtain the hopping terms for the DMFT calculation). We then extract the near Fermi surface $t_{2g}$ derived bands via a fit to  maximally localized Wannier Functions (MLWF)~\cite{PhysRevB.56.12847,PhysRevB.65.035109} derived from $t_{2g}$ orbitals of Ru atom using the  {\sc wannier90} code~\cite{Mostofi2008685} with an energy window from $-3$~eV to $1$~eV. For the cubic Ba material the application is straightforward. For the GdFeO$_3$-distorted Ca and Sr materials we follow the procedure outlined in Ref.~\onlinecite{PhysRevB.91.195149} to find a Wannier basis adapted to the local orientation of each octahedron.   

Figure~\ref{fig:BaRuO3bands} shows the near Fermi surface band structure of BaRuO$_3$ and the Wannier fit to the $t_{2g}$ symmetry states. The Wannier and VASP bands are almost indistinguishable. We observe that the $t_{2g}$ bands identified by the Wannier procedure overlap slightly in energy with other bands both at the lower end of the $t_{2g}$ bands ($E\approx-2.5$~eV) and very near the Fermi level.  The overlap issue is much less severe in the Sr system and does not occur at all in the Ca system~\cite{PhysRevB.91.195149} because the GdFeO$_3$ distortion in those compounds leads to narrower bands that are better separated in energy. The cubic structure of BaRuO$_3$ means that straightforward symmetry considerations enable us to distinguish the $t_{2g}$ bands from the other states.  At energy $E\approx-2.5$~eV, the  overlap is with  oxygen $p$-derived bands. The overlap occurs near the zone center [$\Gamma$ point - denoted by $\mathrm{G}$ in Fig.~\ref{fig:BaRuO3bands}(a)] where there is a sharp symmetry distinction between the states so identification of bands is unambiguous. The other states near and above the Fermi level are determined by a five band Wannier analysis (not shown)  to be of  Ru $e_g$ origin.  Inclusion of beyond-band theory interactions will increase the crystal field splittings, pushing these $e_g$-derived bands farther from near Fermi surface region of interest. We neglect the $e_g$-derived bands henceforth. 

\subsection{Many-body physics}

To treat the many-body physics  of BaRuO$_3$ we  follow Ref.~\onlinecite{PhysRevB.91.195149} and use the one-shot density functional plus dynamical mean field method, in which an effective Hamiltonian for the frontier $t_{2g}$-derived bands is defined as 

\begin{equation}\label{eq:Hamiltonian}
H = H_{\mathrm{kin}} + H_{\mathrm{onsite}},
\end{equation} 
with $H_{\mathrm{kin}}$ obtained by projecting the DFT Hamiltonian onto the Wannier bands discussed above and setting the chemical potential  to ensure that these bands contain four electrons per Ru.

As usual in studies of transition metal oxides, the interaction Hamiltonian  is taken to be site-local and to have  the rotationally invariant Slater-Kanamori form \cite{Imada98}. We use the version appropriate \cite{Georges13} for intra-$t_{2g}$ orbitals, since these are the primary focus of this work
\begin{equation}\label{eq:onsite_SlaterKanamori}
\begin{split}
H_{onsite} & = U\sum_{\alpha}n_{\alpha\uparrow}n_{\alpha\downarrow}  + (U-2J)\sum_{\alpha\neq\beta} n_{\alpha\uparrow}n_{\beta\downarrow} + \\
& + (U-3J)\sum_{\alpha > \beta,\sigma}n_{\alpha\sigma}n_{\beta\sigma} + \\
& + J\sum_{\alpha\neq\beta} ( c^\dagger_{\alpha\uparrow}c^\dagger_{\beta\downarrow}c_{\alpha\downarrow}c_{\beta\uparrow}
+ c^\dagger_{\alpha\uparrow}c^\dagger_{\alpha\downarrow}c_{\beta\downarrow}c_{\beta\uparrow}
),
\end{split}
\end{equation}
where $ \alpha,\beta $ are orbital indexes and $ \sigma $ is the spin index. We set $U=2.3$~eV and $J=0.35$~eV as proposed for the Ca and Sr materials in Ref.~\onlinecite{PhysRevB.91.195149} and solved the impurity model using the hybridization expansion variant of the continuous-time quantum Monte Carlo (CT-HYB) solver as implemented in the  Toolbox for Research on Interacting Quantum Systems (TRIQS) library~\cite{Parcollet2015398,Seth15}.

The momentum integral needed to obtain the on-site Green's function for the DMFT loop is via Gaussian quadrature using  $30^3$ $k$ points for BaRuO$_3$ and $26^3$ $k$ points for Sr- and CaRuO$_3$; the number of k-points is chosen to be  large enough to capture the main features of the density of states.

\section{Results \label{sec:results}}

\begin{figure}[t]
\includegraphics[width=\columnwidth]{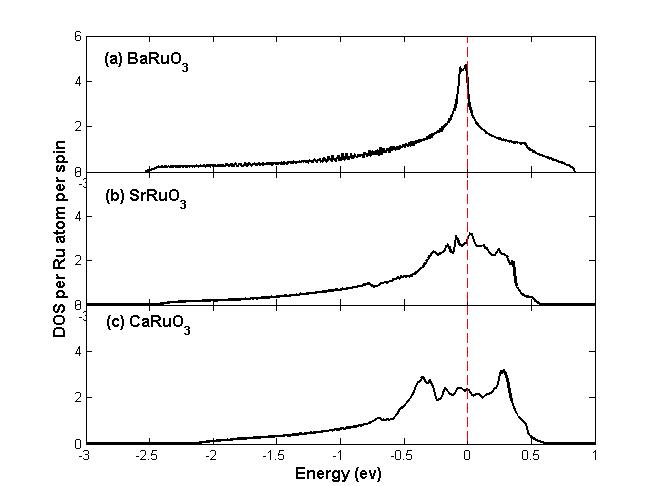}
\caption{(Color online) $t_{2g}$ projected near Fermi surface density of states for BaRuO$_3$, SrRuO$_3$ and CaRuO$_3$, obtained from Wannier fit to calculated band structure. Vertical red dashed line indicates the Fermi level.}
\label{fig:DOScompare}
\end{figure}

\subsection{Density of states}
Figure~\ref{fig:DOScompare} shows the density of states of the three materials, projected onto the Wannier functions corresponding to the  Ru $t_{2g}$ orbitals of interest here. We see that the Ba compound has the largest bandwidth ($\approx 3.6$~eV) and exhibits a near Fermi level  density of states peak arising from a van Hove singularity.  The GdFeO$_3$ distortion reduces the bandwidth and by splitting the van Hove singularity reduces the near Fermi level DOS. The Sr compound has bandwidth of $\approx 3.0$~eV. The Ca material has a larger amplitude GdFeO$_3$ distortion and a correspondingly smaller bandwidth ($\approx 2.6$~eV) and  larger splitting of the van Hove peak. 

\subsection{Magnetic transition temperature}

\begin{figure}[t]
\centering
\includegraphics[width=\columnwidth]{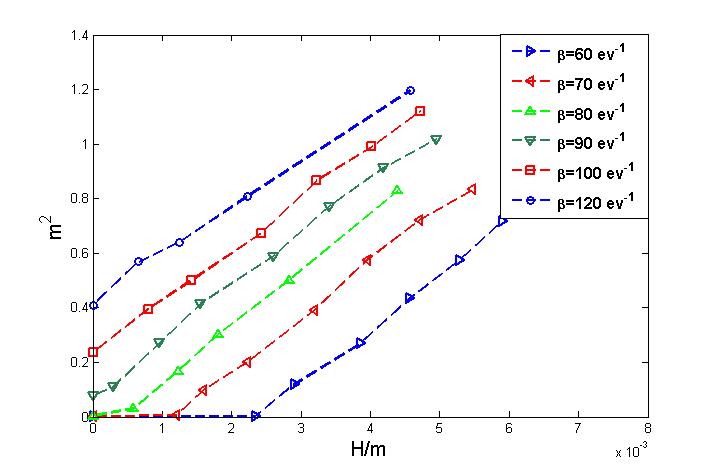}\\
\footnotesize(a)
\includegraphics[width=\columnwidth]{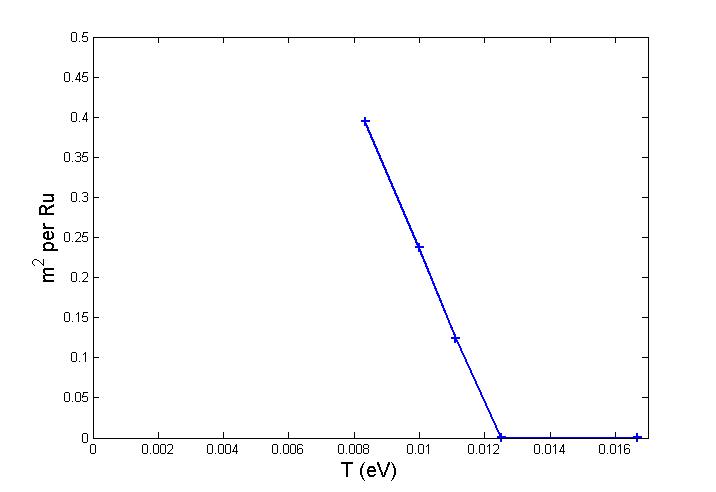}\\
\footnotesize(b)
\caption{(Color online) (a) Square of calculated magnetization $m^2$ of BaRuO$_3$ plotted against ratio of applied field H divided by magnetization $m$ at $U=2.3$~eV, $J=0.35$~eV and temperatures indicated. (b) $m^2$ calculated at $H=0$ plotted against temperature.}
\label{fig:arrott}
\end{figure}

To determine the magnetic transition temperatures we applied small fields $H$ to the Ru $t_{2g}$ orbitals, measured the resulting $t_{2g}$ spin polarization $m$ in the converged DMFT solution,  and plotted $m^2$ against $H/m$ for different $H$ and temperature $T$.  We find that our calculated $m$ fits the Arrott form~\cite{Arrott57}
\begin{equation}
m^2=\frac{1}{B}\frac{H}{m}-\frac{A}{B}(T-T_c)
\end{equation}
very well, and the temperature at which the extrapolated $H/m=0$ value of $m^2$ vanishes provides a good estimate of the transition temperature. To confirm the result we extended the DMFT solution to the ferromagnetic phase and plotted $m^2$ against temperature. The form of the Arrott plots and the agreement between these and the value calculated from the $m^2$ vs $T$ plot also confirms that the transition is second order. In the dynamical mean field approximation used here the transition is mean field, experimental measurements on BaRuO$_3$ reveal non-mean-field exponents \cite{PhysRevLett.101.077206} indicating the importance of fluctuations which would lower the transition temperature relative to the mean field estimate.

Results are shown in Fig.~\ref{fig:arrott} and confirm a transition temperature for BaRuO$_3$ of $T_c\approx 0.012~\mathrm{eV}\sim 140$~K.  This transition temperature is to be compared to the calculated value $T_c\approx 0.017~\mathrm{eV} \sim 200$~K for SrRuO$_3$ and the absence of ferromagnetism in CaRuO$_3$ obtained using the same methods and the same interaction parameters \cite{PhysRevB.91.195149}.  Bearing in mind that mean field approximations such as DMFT overestimate transition temperatures, we consider that the findings are in good agreement with experimental results on this family of materials where CaRuO$_3$ is not magnetically ordered to the lowest temperatures studied, SrRuO$_3$ has a Curie temperature $T_c\approx 160$~K \cite{Cao97} and BaRuO$_3$ has $T_c\approx 60$~K \cite{Jin20052008}. 

\subsection{Self Energies}

\begin{figure}[t]
\includegraphics[width=\columnwidth]{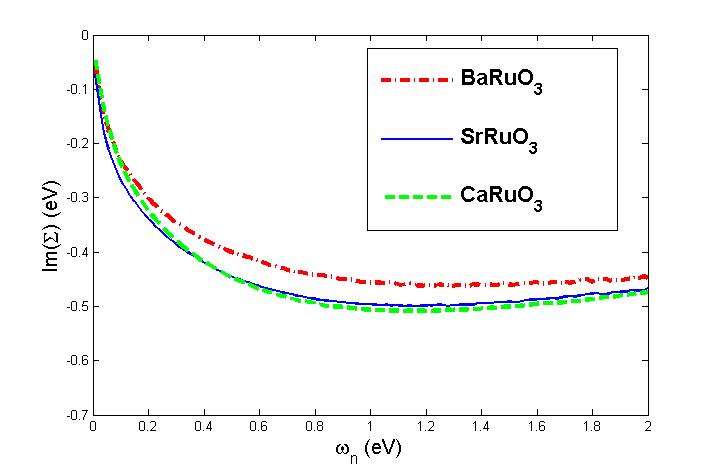}\\(a)\\
\includegraphics[width=\columnwidth]{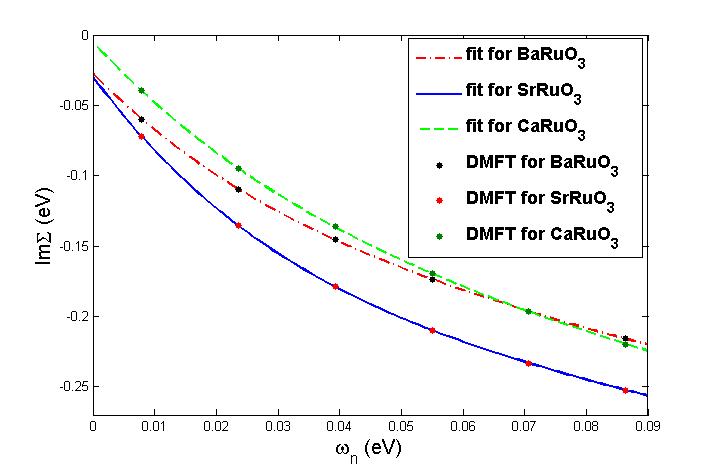}\\(b)
\caption{(Color online) (a) Imaginary part of orbitally averaged self energy of CaRuO$_3$ (dashed lines, green on-line), SrRuO$_3$ (solid line, blue on-line) and BaRuO$_3$ (dash-dotted line, red on-line) calculated in the paramagnetic state at $T=0.0025$~eV  with $U=2.3$~eV and $J=0.35$~eV. (b) Expanded view of low frequency region. The dots  are the DMFT results and the curves are from the fourth order polynomial fit of the last six points of $Im\Sigma(i\omega_n)$.}
\label{fig:selfenergies}
\end{figure}

To better understand the differences in physics between the compound we present in Fig.~\ref{fig:selfenergies} the imaginary part of the Matsubara self energies for the three compounds, calculated using the interaction parameters given above at temperature $T=0.0025$~eV. In the Sr and Ca materials the octahedral rotations lead to small differences between the self energies corresponding to different local orbitals. As the differences between orbitals are not of interest here we present results averaged over all three orbitals. We further restrict our calculations to paramagnetic phases, because the onset of ferromagnetism dramatically changes the self energies. From Fig.~\ref{fig:selfenergies}(a) we see that for $\omega_n > 0.5$~eV, the Ca compound (smallest bandwidth) has the largest magnitude of the self energy and the Ba compound (largest bandwidth) has the smallest. This variation between compounds is consistent with the ``Mott'' picture in which the key parameter is the ratio of an interaction strength to a bandwidth. However  we see from panel b of  Fig.~\ref{fig:selfenergies} that  at low frequency the curves  cross. For $\omega \lesssim0.4$~eV the self energy for the Sr compound becomes larger in magnitude than that for the Ca compound while for  $\omega \lesssim0.07$~eV the self energies for the Ca and Ba materials cross. We expect that at even lower temperatures the self energies for the Ba and Sr materials will cross. This behavior suggests that the very low frequency and temperature limits of the self energy are controlled by the near Fermi level density of states, which is largest for the Ba material and smallest for the Ca material, rather than by the bandwidth~\cite{PhysRevLett.106.096401}.

\section{Analysis \label{sec:analysis}}

In this section we analyze the calculated  variation of transition temperature across the material families. We begin our analysis by considering the Stoner (Hartree-Fock) criterion for magnetism. In its simplest form \cite{doi:10.1080/14786440208564203}, the Stoner criterion relates the onset of magnetism to the product of an interaction and the Fermi surface density of states. Assuming an orbital-independent magnetization $m=\sum_{\alpha}\left(n_{\alpha\uparrow}-n_{\alpha\downarrow}\right)/3$, we find that the change in interaction energy (expectation value of $H_{onsite}$,  Eq.~\eqref{eq:onsite_SlaterKanamori}) is   
\begin{eqnarray}
\delta E_{interaction}=-3(U+2J)\left(\dfrac{m}{2}\right)^2.
\end{eqnarray}
For small $m$, the kinetic energy cost is
\begin{equation}
\delta E_{kinetic}= 3\dfrac{m^2}{4N_0},
\end{equation}
where $N_0$ is the density of states per orbital per spin, averaged over all orbitals. Thus the Stoner criterion for the multi-orbital situation considered here is
\begin{equation}
(U+2J)N_0>1.
\label{eq:Stoner}
\end{equation}

Inspection of Fig.~\ref{fig:DOScompare} shows that the values of $N_0$ are $\sim  1.2$~$~\mathrm{eV}^{-1}$, $\sim  0.97~\mathrm{eV}^{-1}$ and $\sim 0.78~\mathrm{eV}^{-1}$ for BaRuO$_3$, SrRuO$_3$ and CaRuO$_3$ respectively. The Stoner criterion therefore indicates, in clear contradiction to experiment and to our calculated results,  that all three materials should be ferromagnetic, and that the ferromagnetism should be strongest in the Ba material.  This discrepancy suggests that beyond mean-field many-body effects may be important.

\begin{figure}[t]
\centering
\includegraphics[width=\columnwidth]{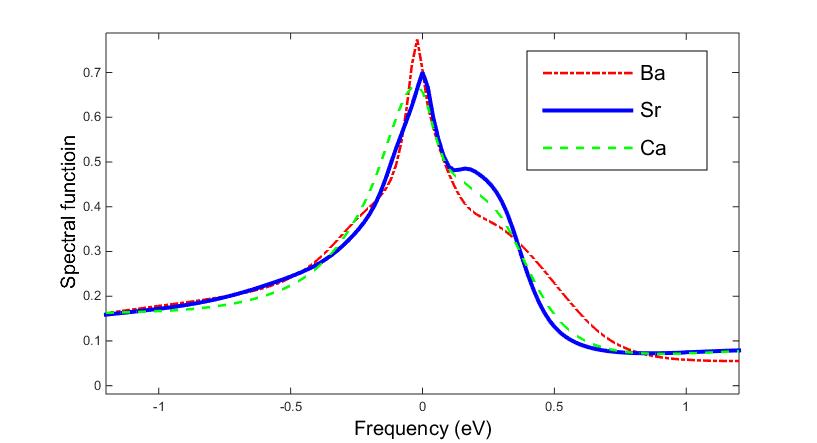}
\caption{(Color online) Momentum-integrated orbitally averaged electron spectral function computed as described in the text  at temperature $T=0.025$~eV. The short-dashed line  (red on-line) is for BaRuO$_3$, the solid line (blue on-line) is for SrRuO$_3$ and the  long-dash line (green on-line) is for CaRuO$_3$. \label{fig:ManybodyDOS}}
\end{figure}   

One possibility is that inelastic scattering broadens the density of states peak. We present in Fig.~\ref{fig:ManybodyDOS} the local spectral function (many-body density of states) $A(\omega)$ calculated by using maximum entropy methods to analytically continue the self energy \cite{PhysRevB.80.045101}  and then inserting the result into the Green function, via
\begin{equation}
  A(\omega) =Im~ \int d^3k ~ Tr\left[\omega + \mu - \hat{H}_{kin}(k) - \hat{\Sigma}(\omega)\right]^{-1} 
\label{A}
\end{equation}
We see that many-body effects substantially reduce the Fermi level density of states of the Ba and Sr materials and slightly reduce that of the Ca material. However, even if we use the many-body density of states, all three materials are  predicted by the Stoner criterion to be ferromagnetic and the Ba material is still predicted to have the strongest magnetism.  We therefore conclude that some parameter other than the value of the Fermi-level density of states is important. A possible explanation was suggested by  Kanamori~\cite{Kanamori63} and investigated in detail for a single band Hubbard model by Vollhardt, Ulmke and coworkers \cite{Ulmke98,Wahle98,Vollhardt01}, and later by  two of us  in the context of vanadate perovskites \cite{Dang13}.  A key issue identified by this work was a strongly skewed density of states, with a peak close to a band edge. In the work of Refs.  \cite{Ulmke98,Wahle98,Vollhardt01,Dang13}, less than half-filled bands were considered, and ferromagnetism was strongest if the Fermi level and density of states peak were close to the lower band edge. In the present situation the band is more than half filled and we expect that ferromagnetism would be strongest if the peak were close to the upper band edge. 

\begin{figure}[t]
\includegraphics[width=\columnwidth]{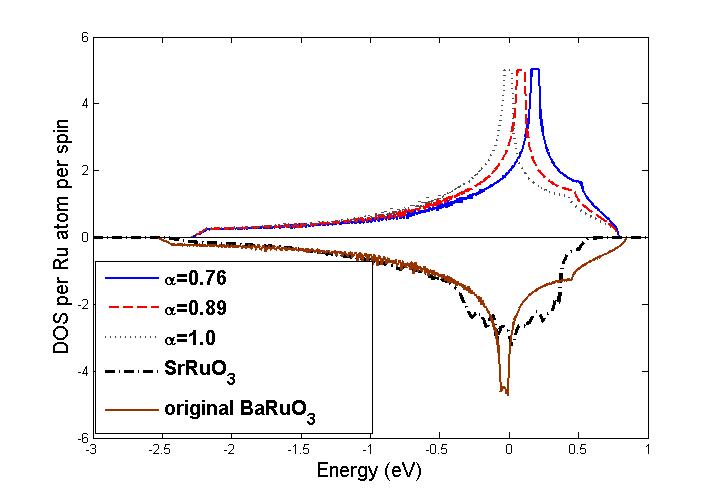}\\(a)\\
\includegraphics[width=\columnwidth]{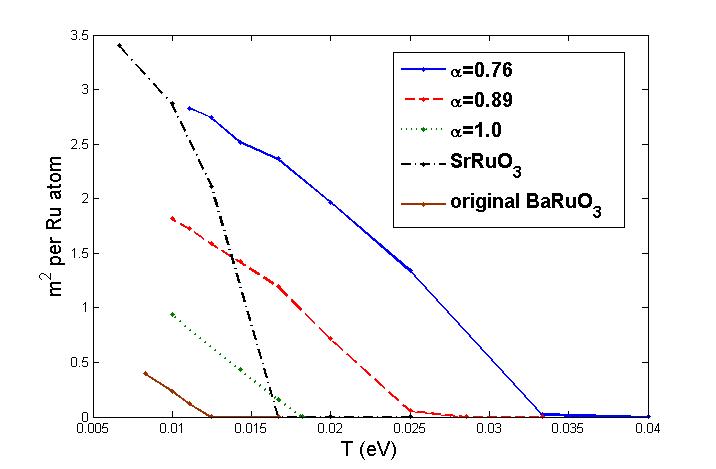}\\(b)
\caption{(Color online) \label{fig:design_dos}(a) The plot of noninteracting density of states: positive half panel: BaRuO$_3$ DOS with bandwidth rescaled to be the same as that of SrRuO$_3$ and DOS peak position shifted towards the upper bandedge. The Fermi level is located at the DOS peak position. The case $\alpha=1$ corresponds to the BaRuO$_3$ DOS with bandwidth scaled to the SrRuO$_3$. The negative half panel: original DOS for BaRuO$_3$ and SrRuO$_3$. (b) $m^2$ vs. $T$ for three typical designed DOS, the original BaRuO$_3$ and SrRuO$_3$.}
\end{figure}

To investigate this possibility we constructed a family of model system densities of states, all of which have the same bandwidth as SrRuO$_3$ but with the van Hove peak at the Fermi level as in BaRuO$_3$. The densities of states differ in the  positions of the upper band edge $E_U$ relative to the Fermi level $E_F$ which we label by $\alpha=(E_U-E_F)/(E^0_U-E^0_F)$, where the superscript $0$ indicates the values for BaRuO$_3$ with only the bandwidth rescaled to the SrRuO$_3$. Several members of this family are shown in Fig.~\ref{fig:design_dos}(a) (note the Fermi level is always at the DOS peak).

For each of these systems we solved the DMFT equations and computed the transition temperatures, finding $T_c\sim 0.033$~eV for $\alpha=0.76$, $\sim 0.025$~eV for $\alpha=0.86$, $\sim 0.018$~eV for $\alpha=1.0$ which should be compared to $T_c\sim 0.0125$~eV for  BaRuO$_3$. The results show that simply rescaling the bandwidth of BaRuO$_3$ ($\alpha=1.0$) increases the  $T_c$ to the value $T_c\sim 0.018$~eV calculated for SrRuO$_3$. A further increase occurs if the  the DOS peak is moved towards the upper band edge. Moving the DOS peak closer to the upper band edge also increases the calculated magnetization  [see Fig.~\ref{fig:design_dos}(b)]. 

We therefore conclude that ferromagnetic transition temperature is controlled by three factors: the DOS at the Fermi level (Stoner theory \cite{doi:10.1080/14786440208564203}), the DOS peak position with respect to the bandedge (Kanamori, Vollhardt and others \cite{Kanamori63,Vollhardt01,Wahle98,Ulmke98,Dang13}), and the bandwidth. CaRuO$_3$ has no ferromagnetism because the strong lattice distortion leads to the splitting of the DOS peak and thus results in small magnitude of DOS at the Fermi level. BaRuO$_3$, despite of a large DOS peak, has larger bandwidth and the DOS peak position farther from the upper bandedge than SrRuO$_3$, explaining the higher $T_c$ of SrRuO$_3$. As seen in Fig.~\ref{fig:design_dos}, adjusting the DOS shape of BaRuO$_3$ to have similar bandwidth and DOS peak position as SrRuO$_3$ will give the Curie temperature much larger than that of SrRuO$_3$.

\section{Self Energies and Correlation Strength \label{sec:correlation}}

The electron correlation strength is a generally important issue for electronically active materials, and the issue is of particular significance in the theory of Hund's metals, where one may consider both the ratio of an interaction parameter  to the bandwidth, and the ratio of an interaction parameter to the near Fermi level density of states \cite{Georges13,PhysRevB.91.195149}.  BaRuO$_3$ highlights this issue, as this material has both the largest bandwidth and the largest Fermi level density of states. 

The correlation strength may  be parametrized by the value of the imaginary part of the self energy. From Fig.~\ref{fig:selfenergies} we see that over the broad energy range, the Ca material has the largest self energy magnitude as expected from its smallest bandwidth. In the low energy range, the self energy curves cross and SrRuO$_3$ has the largest self energy magnitude. At lower temperatures we expect that the low frequency self energies of the BaRuO$_3$ and SrRuO$_3$ would cross and BaRuO$_3$ self energy become the largest.

\begin{table}[b]
\centering
\caption{Intercept $s_0$ and slope $s_1$ obtained from fourth order fit to orbitally averaged $Im\Sigma$ computed at $T=0.0025$~eV and the relative difference $\Delta$ in percentage of the slope $s_1$ obtained from the fitting and from the lowest two Matsubara points. $\Delta$ is defined as the difference between the two slope values divided by $s_1$ value from the fitting.}
\begin{ruledtabular}
\begin{tabular}{c c c c c}
          & $s_0$    & $s_1$  & $s_1 \pi T$ & $\Delta$ \\
\hline
BaRuO$_3$ & -0.02734 & -4.508 & -0.0354     & 30\% \\
SrRuO$_3$ & -0.02974 & -5.929 & -0.0466     & 32\% \\
CaRuO$_3$ & -0.00508 & -4.765 & -0.0374     & 26\% \\
\end{tabular}
\end{ruledtabular}
\label{table:fit}
\end{table}

To analyse the self energy in more detail we fit the lowest six Matsubara points to a fourth order polynomial 
\begin{equation}
Im\Sigma(\omega_n) =\sum_{p=0}^4s_p\omega_n^p .
\label{eq:polyfit}
\end{equation}
where $s_0$ is the residual scattering rate and $s_1$ is an  estimate for $Re\left[d\Sigma/d\omega\right]|_{\omega_n\rightarrow 0}$ which, within the single-site DMFT approximation, yields the mass enhancement via 
\begin{equation}
\frac{m^*}{m}\approx 1-\frac{d Im(\Sigma(i\omega_n))}{d \omega_n}|_{\omega_n\rightarrow 0}\approx 1-s_1.
\label{mstarapprox}
\end{equation}

Two tests of whether the system is in the Fermi liquid regime (so that $1-s_1$ provides a good approximation to the mass enhancement)  are that $s_0\ll Im \Sigma(\omega_n=\pi T)\approx s_1\pi T$ and that the slope defined from the lowest two Matsubara points is in good agreement with the slope defined from the higher order polynomial fit. The low frequency data and the fitted curves are shown in  Fig.~\ref{fig:selfenergies}(b). Table~\ref{table:fit} shows the first two coefficients along with the percent difference between $s_1$ and  the slope defined from the lowest two Matsubara points. We see that for all materials the slopes computed in two different ways agree at the 25-30$\%$ level, indicating that the calculations have at least reached the edge of the Fermi liquid regime. However, for  the Sr and Ba materials the intercept (residual scattering rate) is still about 50\% of the value at the lowest Matsubara frequency, suggesting that these compounds have not quite reached the Fermi liquid regime, so the properties would evolve further as the temperature is lowered.

\begin{figure}
 \includegraphics[width=\columnwidth]{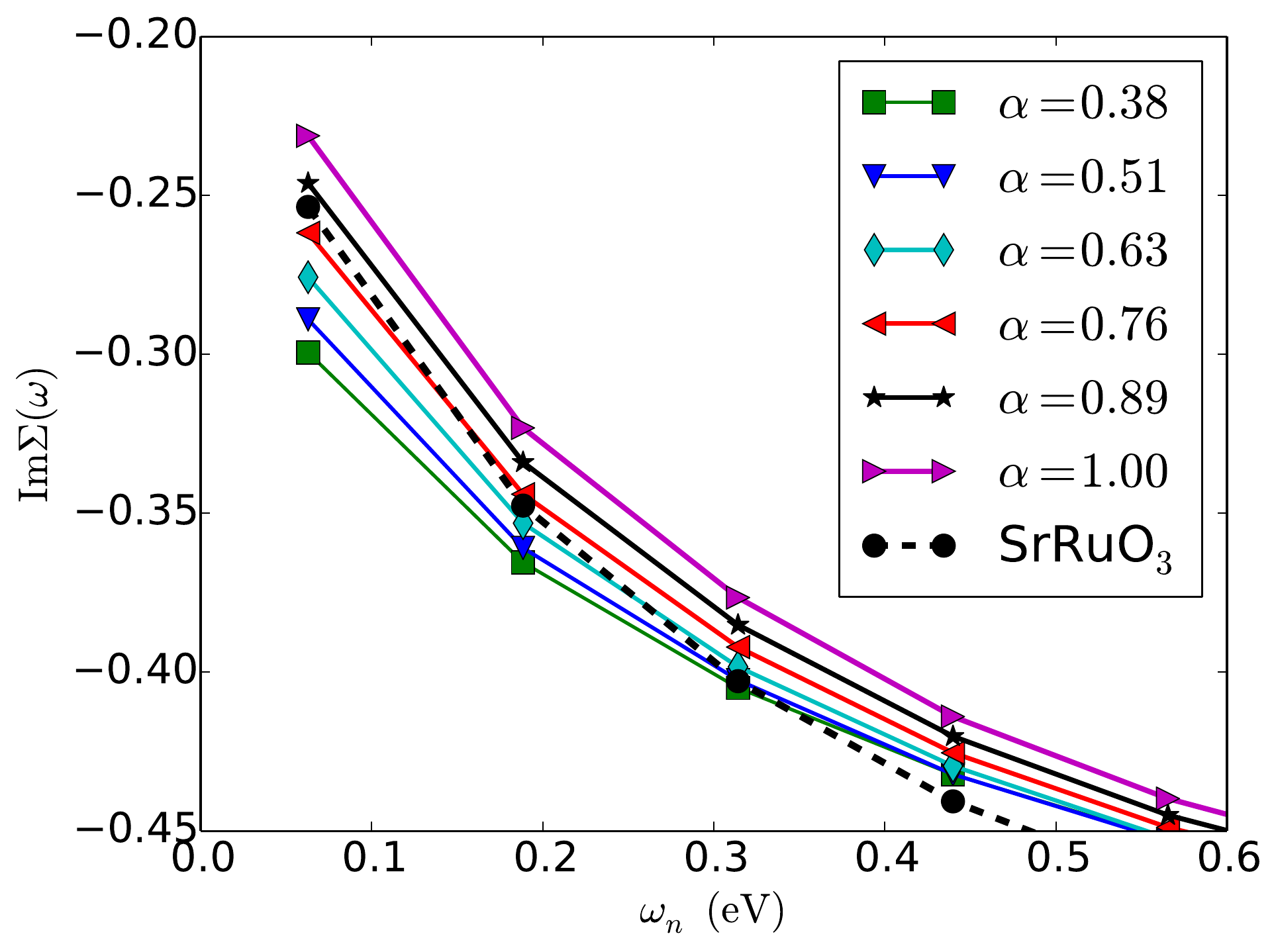}
\caption{\label{fig:self_energy_model} (Color online) Imaginary part of the self energy for model calculation with DOS peak position (represented by the dimensionless parameter $\alpha$) varying. SrRuO$_3$ self energy (circle dashed black line) is also showed for reference. Temperature is fixed at $T=0.02$~eV. Results are obtained for paramagnetic order.}
\end{figure}

At $T=0.0025$~eV the mass enhancement of SrRuO$_3$ is about 6.9 while for CaRuO$_3$ and BaRuO$_3$ it is about 5.5, although as noted for the Sr and Ba compounds the mass enhancement may evolve further as the temperature is lowered. The differences we see between the wide-range and low frequency self energies are consistent with previous work and suggest (in agreement with previous work) that in Hund's metals the low frequency mass enhancement is more sensitive to the Fermi level density of states than to the overall bandwidth~\cite{PhysRevLett.106.096401}. However, the  BaRuO$_3$ results  show that one needs to go to extremely low temperatures before the density of states effect dominates. 

To gain more insight into the relative importance of different factors in the density of states we have computed the mass enhancements for the model  DOS shown in Fig.~\ref{fig:design_dos}(a). Figure~\ref{fig:self_energy_model} shows the results for self energy at $T=0.02$~eV (computed in the paramagnetic state). There is a systematic increase in correlation strength at low frequency as $\alpha$ decreases, implying a stronger Hund's effect when the DOS peak position gets closer to the upper band edge. This increase in the effective mass is consistent with the increase in $T_c$ and suggests that the importance of the Hund's coupling is determined not only by the Fermi level density of states but also by the breaking of particle-hole symmetry.

\section{Conclusion \label{sec:conclusion}}
In this paper we have presented single-site dynamical mean field calculations of the ferromagnetic transition temperature and electronic self energy of cubic BaRuO$_3$. We used interaction parameters $U=2.3$~eV and $J=0.35$~eV obtained for CaRuO$_3$ and SrRuO$_3$ in previous work \cite{PhysRevB.91.195149} which place the material far from the Mott insulating regime and firmly in the Hund's metal regime. We  compared the results to those previously obtained on GdFeO$_3$-distorted SrRuO$_3$ and CaRuO$_3$. In good qualitative agreement with experiment, the calculated ferromagnetic transition temperature of BaRuO$_3$ is positive, but substantially lower than that of SrRuO$_3$.  This agreement provides strong evidence that the single-site dynamical mean field approximation is a reasonable description of the ruthenate materials and suggests that the interaction parameter regime identified for the Sr and Ca materials is correct. 

A very recent theoretical study using an almost identical formulation of the DFT+DMFT methodology studied BaRuO$_3$, considering a range of $U$ and $J$ values and focussing on the self energy and local susceptibility in the paramagnetic phase~\cite{arXiv:1511.01371}. Where parameter values overlap, the results of Ref.~\onlinecite{arXiv:1511.01371} for the self energy are in agreement with those presented here. These authors argued, on the basis of a comparison to the fluctuating moment measured at high temperatures, that a value $J=0.5$~eV is more appropriate than the $J=0.35$~eV considered here. This issue deserves further investigation, but we note that according to Ref.~\onlinecite{PhysRevB.91.195149} this value of $J$ would predict  a ferromagnetic ground state for CaRuO$_3$. The other work~\cite{PhysRevB.87.165139} studied BaRuO$_3$ using a slightly different implementation of DFT+DMFT in which the correlated orbitals were defined as atomic-like $d$ orbitals and the model was defined over a much wider energy range. The parameters chosen in this study were such as to lead to weaker correlation effects (self energies  smaller by a factor of $\sim 2$  than those found here). Extending our analysis of trends across material families to a wider range of parameters and to the wide-band implementation of DFT+DMFT are important issues that might be addressed in future work.

The relation between the non-interacting (band theoretic) density of states and many-body materials  properties is a fundamental and important question in condensed matter physics. A striking feature of the band theory density of states of BaRuO$_3$ is a strong van Hove peak very close to the Fermi level. We find that proximity of the van Hove peak to the Fermi level does not by itself drive dramatic correlation effects at the temperature and energy scales accessible to us. The non-monotonic variation of transition temperature with GdFeO$_3$ rotation amplitude indicates that important features of the magnetism are controlled by features beyond the value of the Fermi-level density of states, in particular the bandwidth and distance from the DOS peak to the upper band edge. This finding is in agreement with previous work~\cite{Kanamori63,Ulmke98,Wahle98,Vollhardt01,Dang13}. We also see that even at the lowest frequency the magnitude of the self energy of BaRuO$_3$ is less than that of SrRuO$_3$, indicating that Hund's metal correlations also are sensitive not only to the Fermi level density of states but also to additional structures  in the density of states farther from the Fermi surface.

\begin{acknowledgments}
H.T.D. acknowledges support from the Deutsche Forschungsgemeinschaft (DFG) within projects FOR 1807 and RTG 1995, as well as the allocation of computing time at J\"ulich Supercomputing Centre and RWTH Aachen University through JARA-HPC. We acknowledge Jernej Mravlje and Antoine Georges for helpful discussions. A.J.M. and Q.H. were supported by  NSF-DMR-01308236 . A.J.M. acknowledges the warm hospitality and stimulating intellectual atmosphere of  the College de France during the preparation of this manuscript.
\end{acknowledgments}

\bibliography{bibofBaRuO3}

\begin{thebibliography}{28}%
\makeatletter
\providecommand \@ifxundefined [1]{%
 \@ifx{#1\undefined}
}%
\providecommand \@ifnum [1]{%
 \ifnum #1\expandafter \@firstoftwo
 \else \expandafter \@secondoftwo
 \fi
}%
\providecommand \@ifx [1]{%
 \ifx #1\expandafter \@firstoftwo
 \else \expandafter \@secondoftwo
 \fi
}%
\providecommand \natexlab [1]{#1}%
\providecommand \enquote  [1]{``#1''}%
\providecommand \bibnamefont  [1]{#1}%
\providecommand \bibfnamefont [1]{#1}%
\providecommand \citenamefont [1]{#1}%
\providecommand \href@noop [0]{\@secondoftwo}%
\providecommand \href [0]{\begingroup \@sanitize@url \@href}%
\providecommand \@href[1]{\@@startlink{#1}\@@href}%
\providecommand \@@href[1]{\endgroup#1\@@endlink}%
\providecommand \@sanitize@url [0]{\catcode `\\12\catcode `\$12\catcode
  `\&12\catcode `\#12\catcode `\^12\catcode `\_12\catcode `\%12\relax}%
\providecommand \@@startlink[1]{}%
\providecommand \@@endlink[0]{}%
\providecommand \url  [0]{\begingroup\@sanitize@url \@url }%
\providecommand \@url [1]{\endgroup\@href {#1}{\urlprefix }}%
\providecommand \urlprefix  [0]{URL }%
\providecommand \Eprint [0]{\href }%
\providecommand \doibase [0]{http://dx.doi.org/}%
\providecommand \selectlanguage [0]{\@gobble}%
\providecommand \bibinfo  [0]{\@secondoftwo}%
\providecommand \bibfield  [0]{\@secondoftwo}%
\providecommand \translation [1]{[#1]}%
\providecommand \BibitemOpen [0]{}%
\providecommand \bibitemStop [0]{}%
\providecommand \bibitemNoStop [0]{.\EOS\space}%
\providecommand \EOS [0]{\spacefactor3000\relax}%
\providecommand \BibitemShut  [1]{\csname bibitem#1\endcsname}%
\let\auto@bib@innerbib\@empty
\bibitem [{\citenamefont {Jin}\ \emph {et~al.}(2008)\citenamefont {Jin},
  \citenamefont {Zhou}, \citenamefont {Goodenough}, \citenamefont {Liu},
  \citenamefont {Zhao}, \citenamefont {Yang}, \citenamefont {Yu}, \citenamefont
  {Yu}, \citenamefont {Katsura}, \citenamefont {Shatskiy},\ and\ \citenamefont
  {Ito}}]{Jin20052008}%
  \BibitemOpen
  \bibfield  {author} {\bibinfo {author} {\bibfnamefont {C.-Q.}\ \bibnamefont
  {Jin}}, \bibinfo {author} {\bibfnamefont {J.-S.}\ \bibnamefont {Zhou}},
  \bibinfo {author} {\bibfnamefont {J.~B.}\ \bibnamefont {Goodenough}},
  \bibinfo {author} {\bibfnamefont {Q.~Q.}\ \bibnamefont {Liu}}, \bibinfo
  {author} {\bibfnamefont {J.~G.}\ \bibnamefont {Zhao}}, \bibinfo {author}
  {\bibfnamefont {L.~X.}\ \bibnamefont {Yang}}, \bibinfo {author}
  {\bibfnamefont {Y.}~\bibnamefont {Yu}}, \bibinfo {author} {\bibfnamefont
  {R.~C.}\ \bibnamefont {Yu}}, \bibinfo {author} {\bibfnamefont
  {T.}~\bibnamefont {Katsura}}, \bibinfo {author} {\bibfnamefont
  {A.}~\bibnamefont {Shatskiy}}, \ and\ \bibinfo {author} {\bibfnamefont
  {E.}~\bibnamefont {Ito}},\ }\href {\doibase 10.1073/pnas.0710928105}
  {\bibfield  {journal} {\bibinfo  {journal} {Proceedings of the National
  Academy of Sciences}\ }\textbf {\bibinfo {volume} {105}},\ \bibinfo {pages}
  {7115} (\bibinfo {year} {2008})}\BibitemShut {NoStop}%
\bibitem [{\citenamefont {Zhou}\ \emph {et~al.}(2008)\citenamefont {Zhou},
  \citenamefont {Matsubayashi}, \citenamefont {Uwatoko}, \citenamefont {Jin},
  \citenamefont {Cheng}, \citenamefont {Goodenough}, \citenamefont {Liu},
  \citenamefont {Katsura}, \citenamefont {Shatskiy},\ and\ \citenamefont
  {Ito}}]{PhysRevLett.101.077206}%
  \BibitemOpen
  \bibfield  {author} {\bibinfo {author} {\bibfnamefont {J.-S.}\ \bibnamefont
  {Zhou}}, \bibinfo {author} {\bibfnamefont {K.}~\bibnamefont {Matsubayashi}},
  \bibinfo {author} {\bibfnamefont {Y.}~\bibnamefont {Uwatoko}}, \bibinfo
  {author} {\bibfnamefont {C.-Q.}\ \bibnamefont {Jin}}, \bibinfo {author}
  {\bibfnamefont {J.-G.}\ \bibnamefont {Cheng}}, \bibinfo {author}
  {\bibfnamefont {J.~B.}\ \bibnamefont {Goodenough}}, \bibinfo {author}
  {\bibfnamefont {Q.~Q.}\ \bibnamefont {Liu}}, \bibinfo {author} {\bibfnamefont
  {T.}~\bibnamefont {Katsura}}, \bibinfo {author} {\bibfnamefont
  {A.}~\bibnamefont {Shatskiy}}, \ and\ \bibinfo {author} {\bibfnamefont
  {E.}~\bibnamefont {Ito}},\ }\href {\doibase 10.1103/PhysRevLett.101.077206}
  {\bibfield  {journal} {\bibinfo  {journal} {Phys. Rev. Lett.}\ }\textbf
  {\bibinfo {volume} {101}},\ \bibinfo {pages} {077206} (\bibinfo {year}
  {2008})}\BibitemShut {NoStop}%
\bibitem [{\citenamefont {Dang}\ \emph {et~al.}(2015)\citenamefont {Dang},
  \citenamefont {Mravlje}, \citenamefont {Georges},\ and\ \citenamefont
  {Millis}}]{PhysRevB.91.195149}%
  \BibitemOpen
  \bibfield  {author} {\bibinfo {author} {\bibfnamefont {H.~T.}\ \bibnamefont
  {Dang}}, \bibinfo {author} {\bibfnamefont {J.}~\bibnamefont {Mravlje}},
  \bibinfo {author} {\bibfnamefont {A.}~\bibnamefont {Georges}}, \ and\
  \bibinfo {author} {\bibfnamefont {A.~J.}\ \bibnamefont {Millis}},\ }\href
  {\doibase 10.1103/PhysRevB.91.195149} {\bibfield  {journal} {\bibinfo
  {journal} {Phys. Rev. B}\ }\textbf {\bibinfo {volume} {91}},\ \bibinfo
  {pages} {195149} (\bibinfo {year} {2015})}\BibitemShut {NoStop}%
\bibitem [{\citenamefont {Georges}\ \emph {et~al.}(2013)\citenamefont
  {Georges}, \citenamefont {Medici},\ and\ \citenamefont
  {Mravlje}}]{Georges13}%
  \BibitemOpen
  \bibfield  {author} {\bibinfo {author} {\bibfnamefont {A.}~\bibnamefont
  {Georges}}, \bibinfo {author} {\bibfnamefont {L.~d.}\ \bibnamefont {Medici}},
  \ and\ \bibinfo {author} {\bibfnamefont {J.}~\bibnamefont {Mravlje}},\ }\href
  {\doibase 10.1146/annurev-conmatphys-020911-125045} {\bibfield  {journal}
  {\bibinfo  {journal} {Annual Review of Condensed Matter Physics}\ }\textbf
  {\bibinfo {volume} {4}},\ \bibinfo {pages} {137} (\bibinfo {year}
  {2013})}\BibitemShut {NoStop}%
\bibitem [{\citenamefont {Huang}\ and\ \citenamefont
  {Ao}(2013)}]{PhysRevB.87.165139}%
  \BibitemOpen
  \bibfield  {author} {\bibinfo {author} {\bibfnamefont {L.}~\bibnamefont
  {Huang}}\ and\ \bibinfo {author} {\bibfnamefont {B.}~\bibnamefont {Ao}},\
  }\href {\doibase 10.1103/PhysRevB.87.165139} {\bibfield  {journal} {\bibinfo
  {journal} {Phys. Rev. B}\ }\textbf {\bibinfo {volume} {87}},\ \bibinfo
  {pages} {165139} (\bibinfo {year} {2013})}\BibitemShut {NoStop}%
\bibitem [{\citenamefont {Dasari}\ \emph {et~al.}(2015)\citenamefont {Dasari},
  \citenamefont {Yamijala}, \citenamefont {Jain}, \citenamefont {Dasgupta},
  \citenamefont {Moreno}, \citenamefont {Jarrell},\ and\ \citenamefont
  {Vidhyadhiraja}}]{arXiv:1511.01371}%
  \BibitemOpen
  \bibfield  {author} {\bibinfo {author} {\bibfnamefont {N.}~\bibnamefont
  {Dasari}}, \bibinfo {author} {\bibfnamefont {S.~R. K. C.~S.}\ \bibnamefont
  {Yamijala}}, \bibinfo {author} {\bibfnamefont {M.}~\bibnamefont {Jain}},
  \bibinfo {author} {\bibfnamefont {T.~S.}\ \bibnamefont {Dasgupta}}, \bibinfo
  {author} {\bibfnamefont {J.}~\bibnamefont {Moreno}}, \bibinfo {author}
  {\bibfnamefont {M.}~\bibnamefont {Jarrell}}, \ and\ \bibinfo {author}
  {\bibfnamefont {N.~S.}\ \bibnamefont {Vidhyadhiraja}},\ }\href@noop {} {\
  (\bibinfo {year} {2015})},\ \Eprint {http://arxiv.org/abs/1507.00175}
  {arXiv:1507.00175 [cond-mat.str-el]} \BibitemShut {NoStop}%
\bibitem [{\citenamefont {Jones}\ \emph {et~al.}(1989)\citenamefont {Jones},
  \citenamefont {Battle}, \citenamefont {Lightfoot},\ and\ \citenamefont
  {Harrison}}]{Jones89}%
  \BibitemOpen
  \bibfield  {author} {\bibinfo {author} {\bibfnamefont {C.~W.}\ \bibnamefont
  {Jones}}, \bibinfo {author} {\bibfnamefont {P.~D.}\ \bibnamefont {Battle}},
  \bibinfo {author} {\bibfnamefont {P.}~\bibnamefont {Lightfoot}}, \ and\
  \bibinfo {author} {\bibfnamefont {W.~T.~A.}\ \bibnamefont {Harrison}},\
  }\href {\doibase 10.1107/S0108270188012077} {\bibfield  {journal} {\bibinfo
  {journal} {Acta Crystallographica Section C}\ }\textbf {\bibinfo {volume}
  {45}},\ \bibinfo {pages} {365} (\bibinfo {year} {1989})}\BibitemShut
  {NoStop}%
\bibitem [{\citenamefont {Bensch}\ \emph {et~al.}(1990)\citenamefont {Bensch},
  \citenamefont {Schmalle},\ and\ \citenamefont {Reller}}]{Bensch90}%
  \BibitemOpen
  \bibfield  {author} {\bibinfo {author} {\bibfnamefont {W.}~\bibnamefont
  {Bensch}}, \bibinfo {author} {\bibfnamefont {H.~W.}\ \bibnamefont
  {Schmalle}}, \ and\ \bibinfo {author} {\bibfnamefont {A.}~\bibnamefont
  {Reller}},\ }\href {\doibase 10.1016/0167-2738(90)90481-6} {\bibfield
  {journal} {\bibinfo  {journal} {Solid State Ionics}\ }\textbf {\bibinfo
  {volume} {43}},\ \bibinfo {pages} {171 } (\bibinfo {year}
  {1990})}\BibitemShut {NoStop}%
\bibitem [{\citenamefont {Kresse}\ and\ \citenamefont
  {Hafner}(1993)}]{Kresse93}%
  \BibitemOpen
  \bibfield  {author} {\bibinfo {author} {\bibfnamefont {G.}~\bibnamefont
  {Kresse}}\ and\ \bibinfo {author} {\bibfnamefont {J.}~\bibnamefont
  {Hafner}},\ }\href {\doibase 10.1103/PhysRevB.47.558} {\bibfield  {journal}
  {\bibinfo  {journal} {Phys. Rev. B}\ }\textbf {\bibinfo {volume} {47}},\
  \bibinfo {pages} {558} (\bibinfo {year} {1993})}\BibitemShut {NoStop}%
\bibitem [{\citenamefont {Kresse}\ and\ \citenamefont
  {Furthm\"uller}(1996{\natexlab{a}})}]{Kresse96a}%
  \BibitemOpen
  \bibfield  {author} {\bibinfo {author} {\bibfnamefont {G.}~\bibnamefont
  {Kresse}}\ and\ \bibinfo {author} {\bibfnamefont {J.}~\bibnamefont
  {Furthm\"uller}},\ }\href {\doibase 10.1016/0927-0256(96)00008-0} {\bibfield
  {journal} {\bibinfo  {journal} {Comput. Mat. Sci.}\ }\textbf {\bibinfo
  {volume} {6}},\ \bibinfo {pages} {15} (\bibinfo {year}
  {1996}{\natexlab{a}})}\BibitemShut {NoStop}%
\bibitem [{\citenamefont {Kresse}\ and\ \citenamefont
  {Furthm\"uller}(1996{\natexlab{b}})}]{Kresse96b}%
  \BibitemOpen
  \bibfield  {author} {\bibinfo {author} {\bibfnamefont {G.}~\bibnamefont
  {Kresse}}\ and\ \bibinfo {author} {\bibfnamefont {J.}~\bibnamefont
  {Furthm\"uller}},\ }\href {\doibase 10.1103/PhysRevB.54.11169} {\bibfield
  {journal} {\bibinfo  {journal} {Phys. Rev. B}\ }\textbf {\bibinfo {volume}
  {54}},\ \bibinfo {pages} {11169} (\bibinfo {year}
  {1996}{\natexlab{b}})}\BibitemShut {NoStop}%
\bibitem [{\citenamefont {Kresse}\ and\ \citenamefont
  {Joubert}(1999)}]{Kresse99}%
  \BibitemOpen
  \bibfield  {author} {\bibinfo {author} {\bibfnamefont {G.}~\bibnamefont
  {Kresse}}\ and\ \bibinfo {author} {\bibfnamefont {D.}~\bibnamefont
  {Joubert}},\ }\href {\doibase 10.1103/PhysRevB.59.1758} {\bibfield  {journal}
  {\bibinfo  {journal} {Phys. Rev. B}\ }\textbf {\bibinfo {volume} {59}},\
  \bibinfo {pages} {1758} (\bibinfo {year} {1999})}\BibitemShut {NoStop}%
\bibitem [{\citenamefont {Marzari}\ and\ \citenamefont
  {Vanderbilt}(1997)}]{PhysRevB.56.12847}%
  \BibitemOpen
  \bibfield  {author} {\bibinfo {author} {\bibfnamefont {N.}~\bibnamefont
  {Marzari}}\ and\ \bibinfo {author} {\bibfnamefont {D.}~\bibnamefont
  {Vanderbilt}},\ }\href {\doibase 10.1103/PhysRevB.56.12847} {\bibfield
  {journal} {\bibinfo  {journal} {Phys. Rev. B}\ }\textbf {\bibinfo {volume}
  {56}},\ \bibinfo {pages} {12847} (\bibinfo {year} {1997})}\BibitemShut
  {NoStop}%
\bibitem [{\citenamefont {Souza}\ \emph {et~al.}(2001)\citenamefont {Souza},
  \citenamefont {Marzari},\ and\ \citenamefont
  {Vanderbilt}}]{PhysRevB.65.035109}%
  \BibitemOpen
  \bibfield  {author} {\bibinfo {author} {\bibfnamefont {I.}~\bibnamefont
  {Souza}}, \bibinfo {author} {\bibfnamefont {N.}~\bibnamefont {Marzari}}, \
  and\ \bibinfo {author} {\bibfnamefont {D.}~\bibnamefont {Vanderbilt}},\
  }\href {\doibase 10.1103/PhysRevB.65.035109} {\bibfield  {journal} {\bibinfo
  {journal} {Phys. Rev. B}\ }\textbf {\bibinfo {volume} {65}},\ \bibinfo
  {pages} {035109} (\bibinfo {year} {2001})}\BibitemShut {NoStop}%
\bibitem [{\citenamefont {Mostofi}\ \emph {et~al.}(2008)\citenamefont
  {Mostofi}, \citenamefont {Yates}, \citenamefont {Lee}, \citenamefont {Souza},
  \citenamefont {Vanderbilt},\ and\ \citenamefont {Marzari}}]{Mostofi2008685}%
  \BibitemOpen
  \bibfield  {author} {\bibinfo {author} {\bibfnamefont {A.~A.}\ \bibnamefont
  {Mostofi}}, \bibinfo {author} {\bibfnamefont {J.~R.}\ \bibnamefont {Yates}},
  \bibinfo {author} {\bibfnamefont {Y.-S.}\ \bibnamefont {Lee}}, \bibinfo
  {author} {\bibfnamefont {I.}~\bibnamefont {Souza}}, \bibinfo {author}
  {\bibfnamefont {D.}~\bibnamefont {Vanderbilt}}, \ and\ \bibinfo {author}
  {\bibfnamefont {N.}~\bibnamefont {Marzari}},\ }\href {\doibase
  http://dx.doi.org/10.1016/j.cpc.2007.11.016} {\bibfield  {journal} {\bibinfo
  {journal} {Computer Physics Communications}\ }\textbf {\bibinfo {volume}
  {178}},\ \bibinfo {pages} {685 } (\bibinfo {year} {2008})}\BibitemShut
  {NoStop}%
\bibitem [{\citenamefont {Imada}\ \emph {et~al.}(1998)\citenamefont {Imada},
  \citenamefont {Fujimori},\ and\ \citenamefont {Tokura}}]{Imada98}%
  \BibitemOpen
  \bibfield  {author} {\bibinfo {author} {\bibfnamefont {M.}~\bibnamefont
  {Imada}}, \bibinfo {author} {\bibfnamefont {A.}~\bibnamefont {Fujimori}}, \
  and\ \bibinfo {author} {\bibfnamefont {Y.}~\bibnamefont {Tokura}},\ }\href
  {\doibase 10.1103/RevModPhys.70.1039} {\bibfield  {journal} {\bibinfo
  {journal} {Rev. Mod. Phys.}\ }\textbf {\bibinfo {volume} {70}},\ \bibinfo
  {pages} {1039} (\bibinfo {year} {1998})}\BibitemShut {NoStop}%
\bibitem [{\citenamefont {Parcollet}\ \emph {et~al.}(2015)\citenamefont
  {Parcollet}, \citenamefont {Ferrero}, \citenamefont {Ayral}, \citenamefont
  {Hafermann}, \citenamefont {Krivenko}, \citenamefont {Messio},\ and\
  \citenamefont {Seth}}]{Parcollet2015398}%
  \BibitemOpen
  \bibfield  {author} {\bibinfo {author} {\bibfnamefont {O.}~\bibnamefont
  {Parcollet}}, \bibinfo {author} {\bibfnamefont {M.}~\bibnamefont {Ferrero}},
  \bibinfo {author} {\bibfnamefont {T.}~\bibnamefont {Ayral}}, \bibinfo
  {author} {\bibfnamefont {H.}~\bibnamefont {Hafermann}}, \bibinfo {author}
  {\bibfnamefont {I.}~\bibnamefont {Krivenko}}, \bibinfo {author}
  {\bibfnamefont {L.}~\bibnamefont {Messio}}, \ and\ \bibinfo {author}
  {\bibfnamefont {P.}~\bibnamefont {Seth}},\ }\href {\doibase
  http://dx.doi.org/10.1016/j.cpc.2015.04.023} {\bibfield  {journal} {\bibinfo
  {journal} {Computer Physics Communications}\ }\textbf {\bibinfo {volume}
  {196}},\ \bibinfo {pages} {398 } (\bibinfo {year} {2015})}\BibitemShut
  {NoStop}%
\bibitem [{\citenamefont {Seth}\ \emph {et~al.}(2015)\citenamefont {Seth},
  \citenamefont {Krivenko}, \citenamefont {Ferrero},\ and\ \citenamefont
  {Parcollet}}]{Seth15}%
  \BibitemOpen
  \bibfield  {author} {\bibinfo {author} {\bibfnamefont {P.}~\bibnamefont
  {Seth}}, \bibinfo {author} {\bibfnamefont {I.}~\bibnamefont {Krivenko}},
  \bibinfo {author} {\bibfnamefont {M.}~\bibnamefont {Ferrero}}, \ and\
  \bibinfo {author} {\bibfnamefont {O.}~\bibnamefont {Parcollet}},\ }\href@noop
  {} {\  (\bibinfo {year} {2015})},\ \Eprint {http://arxiv.org/abs/1507.00175}
  {arXiv:1507.00175 [cond-mat.str-el]} \BibitemShut {NoStop}%
\bibitem [{\citenamefont {Arrott}(1957)}]{Arrott57}%
  \BibitemOpen
  \bibfield  {author} {\bibinfo {author} {\bibfnamefont {A.}~\bibnamefont
  {Arrott}},\ }\href {\doibase 10.1103/PhysRev.108.1394} {\bibfield  {journal}
  {\bibinfo  {journal} {Phys. Rev.}\ }\textbf {\bibinfo {volume} {108}},\
  \bibinfo {pages} {1394} (\bibinfo {year} {1957})}\BibitemShut {NoStop}%
\bibitem [{\citenamefont {Cao}\ \emph {et~al.}(1997)\citenamefont {Cao},
  \citenamefont {McCall}, \citenamefont {Shepard}, \citenamefont {Crow},\ and\
  \citenamefont {Guertin}}]{Cao97}%
  \BibitemOpen
  \bibfield  {author} {\bibinfo {author} {\bibfnamefont {G.}~\bibnamefont
  {Cao}}, \bibinfo {author} {\bibfnamefont {S.}~\bibnamefont {McCall}},
  \bibinfo {author} {\bibfnamefont {M.}~\bibnamefont {Shepard}}, \bibinfo
  {author} {\bibfnamefont {J.~E.}\ \bibnamefont {Crow}}, \ and\ \bibinfo
  {author} {\bibfnamefont {R.~P.}\ \bibnamefont {Guertin}},\ }\href {\doibase
  10.1103/PhysRevB.56.321} {\bibfield  {journal} {\bibinfo  {journal} {Phys.
  Rev. B}\ }\textbf {\bibinfo {volume} {56}},\ \bibinfo {pages} {321} (\bibinfo
  {year} {1997})}\BibitemShut {NoStop}%
\bibitem [{\citenamefont {Mravlje}\ \emph {et~al.}(2011)\citenamefont
  {Mravlje}, \citenamefont {Aichhorn}, \citenamefont {Miyake}, \citenamefont
  {Haule}, \citenamefont {Kotliar},\ and\ \citenamefont
  {Georges}}]{PhysRevLett.106.096401}%
  \BibitemOpen
  \bibfield  {author} {\bibinfo {author} {\bibfnamefont {J.}~\bibnamefont
  {Mravlje}}, \bibinfo {author} {\bibfnamefont {M.}~\bibnamefont {Aichhorn}},
  \bibinfo {author} {\bibfnamefont {T.}~\bibnamefont {Miyake}}, \bibinfo
  {author} {\bibfnamefont {K.}~\bibnamefont {Haule}}, \bibinfo {author}
  {\bibfnamefont {G.}~\bibnamefont {Kotliar}}, \ and\ \bibinfo {author}
  {\bibfnamefont {A.}~\bibnamefont {Georges}},\ }\href {\doibase
  10.1103/PhysRevLett.106.096401} {\bibfield  {journal} {\bibinfo  {journal}
  {Phys. Rev. Lett.}\ }\textbf {\bibinfo {volume} {106}},\ \bibinfo {pages}
  {096401} (\bibinfo {year} {2011})}\BibitemShut {NoStop}%
\bibitem [{\citenamefont {Stoner}(1927)}]{doi:10.1080/14786440208564203}%
  \BibitemOpen
  \bibfield  {author} {\bibinfo {author} {\bibfnamefont {E.~C.}\ \bibnamefont
  {Stoner}},\ }\href {\doibase 10.1080/14786440208564203} {\bibfield  {journal}
  {\bibinfo  {journal} {Philosophical Magazine and Journal of Science}\
  }\textbf {\bibinfo {volume} {3}},\ \bibinfo {pages} {336} (\bibinfo {year}
  {1927})}\BibitemShut {NoStop}%
\bibitem [{\citenamefont {Wang}\ \emph {et~al.}(2009)\citenamefont {Wang},
  \citenamefont {Gull}, \citenamefont {de' Medici}, \citenamefont {Capone},\
  and\ \citenamefont {Millis}}]{PhysRevB.80.045101}%
  \BibitemOpen
  \bibfield  {author} {\bibinfo {author} {\bibfnamefont {X.}~\bibnamefont
  {Wang}}, \bibinfo {author} {\bibfnamefont {E.}~\bibnamefont {Gull}}, \bibinfo
  {author} {\bibfnamefont {L.}~\bibnamefont {de' Medici}}, \bibinfo {author}
  {\bibfnamefont {M.}~\bibnamefont {Capone}}, \ and\ \bibinfo {author}
  {\bibfnamefont {A.~J.}\ \bibnamefont {Millis}},\ }\href {\doibase
  10.1103/PhysRevB.80.045101} {\bibfield  {journal} {\bibinfo  {journal} {Phys.
  Rev. B}\ }\textbf {\bibinfo {volume} {80}},\ \bibinfo {pages} {045101}
  (\bibinfo {year} {2009})}\BibitemShut {NoStop}%
\bibitem [{\citenamefont {Kanamori}(1963)}]{Kanamori63}%
  \BibitemOpen
  \bibfield  {author} {\bibinfo {author} {\bibfnamefont {J.}~\bibnamefont
  {Kanamori}},\ }\href {\doibase 10.1143/PTP.30.275} {\bibfield  {journal}
  {\bibinfo  {journal} {Progress of Theoretical Physics}\ }\textbf {\bibinfo
  {volume} {30}},\ \bibinfo {pages} {275} (\bibinfo {year} {1963})}\BibitemShut
  {NoStop}%
\bibitem [{\citenamefont {Ulmke}(1998)}]{Ulmke98}%
  \BibitemOpen
  \bibfield  {author} {\bibinfo {author} {\bibfnamefont {M.}~\bibnamefont
  {Ulmke}},\ }\href {\doibase 10.1007/s100510050186} {\bibfield  {journal}
  {\bibinfo  {journal} {Eur. Phys. J. B}\ }\textbf {\bibinfo {volume} {1}},\
  \bibinfo {pages} {301} (\bibinfo {year} {1998})}\BibitemShut {NoStop}%
\bibitem [{\citenamefont {Wahle}\ \emph {et~al.}(1998)\citenamefont {Wahle},
  \citenamefont {Bl\"umer}, \citenamefont {Schlipf}, \citenamefont {Held},\
  and\ \citenamefont {Vollhardt}}]{Wahle98}%
  \BibitemOpen
  \bibfield  {author} {\bibinfo {author} {\bibfnamefont {J.}~\bibnamefont
  {Wahle}}, \bibinfo {author} {\bibfnamefont {N.}~\bibnamefont {Bl\"umer}},
  \bibinfo {author} {\bibfnamefont {J.}~\bibnamefont {Schlipf}}, \bibinfo
  {author} {\bibfnamefont {K.}~\bibnamefont {Held}}, \ and\ \bibinfo {author}
  {\bibfnamefont {D.}~\bibnamefont {Vollhardt}},\ }\href {\doibase
  10.1103/PhysRevB.58.12749} {\bibfield  {journal} {\bibinfo  {journal} {Phys.
  Rev. B}\ }\textbf {\bibinfo {volume} {58}},\ \bibinfo {pages} {12749}
  (\bibinfo {year} {1998})}\BibitemShut {NoStop}%
\bibitem [{\citenamefont {Vollhardt}\ \emph {et~al.}(2001)\citenamefont
  {Vollhardt}, \citenamefont {Bl\"umer}, \citenamefont {Held},\ and\
  \citenamefont {Kollar}}]{Vollhardt01}%
  \BibitemOpen
  \bibfield  {author} {\bibinfo {author} {\bibfnamefont {D.}~\bibnamefont
  {Vollhardt}}, \bibinfo {author} {\bibfnamefont {N.}~\bibnamefont {Bl\"umer}},
  \bibinfo {author} {\bibfnamefont {K.}~\bibnamefont {Held}}, \ and\ \bibinfo
  {author} {\bibfnamefont {M.}~\bibnamefont {Kollar}},\ }\enquote {\bibinfo
  {title} {Band-ferromagnetism ground-state and finite-temperature
  phenomena},}\ \ (\bibinfo  {publisher} {Springer, New York},\ \bibinfo {year}
  {2001})\ Chap.\ \bibinfo {chapter} {Metallic Ferromagetism - An Electronic
  Correlation Phenomenon}, p.\ \bibinfo {pages} {191},\ \bibinfo {note}
  {\href{http://arxiv.org/abs/cond-mat/0012203}{arXiv:cond-mat/0012203}}\BibitemShut
  {NoStop}%
\bibitem [{\citenamefont {Dang}\ and\ \citenamefont {Millis}(2013)}]{Dang13}%
  \BibitemOpen
  \bibfield  {author} {\bibinfo {author} {\bibfnamefont {H.~T.}\ \bibnamefont
  {Dang}}\ and\ \bibinfo {author} {\bibfnamefont {A.~J.}\ \bibnamefont
  {Millis}},\ }\href {\doibase 10.1103/PhysRevB.87.155127} {\bibfield
  {journal} {\bibinfo  {journal} {Phys. Rev. B}\ }\textbf {\bibinfo {volume}
  {87}},\ \bibinfo {pages} {155127} (\bibinfo {year} {2013})}\BibitemShut
  {NoStop}%
\end{thebibliography}%

\end{document}